\documentclass[twocolumn,twoside,preprintnumbers,amsmath,amssymb,showkeys]{revtex4}
\usepackage{epsfig}
\usepackage{hypernat}
\usepackage{color}
\usepackage{ifpdf}

\newif\ifcomment
\newif\ifprint
\printtrue

\ifpdf
\ifprint
 \usepackage[linkbordercolor={0 0 .8},%
             citebordercolor={.8 0 0},%
             urlbordercolor={.8 0 .8},%
             raiselinks=true,%
             pdfborder={0 0 1 [3]}]{hyperref}
\else
 \usepackage[pdftex,backref,pagebackref,colorlinks,
             linkcolor=blue,citecolor=blue,anchorcolor=blue,
             pagecolor=blue,urlcolor=red,a4paper]{hyperref}
\fi
\pdfoutput=1        
\pdfcatalog{/PageMode(/UseOutlines)} 
\else 
\ifprint
 \usepackage[linkbordercolor={0 0 .8},%
             citebordercolor={.8 0 0},%
             urlbordercolor={.8 0 .8},%
             raiselinks=true,%
             pdfborder={0 0 1 [3]}]{hyperref}
\else
 \usepackage[dvips,backref,pagebackref,colorlinks,
             linkcolor=blue,citecolor=blue,anchorcolor=blue,
             pagecolor=blue,urlcolor=red,a4paper]{hyperref}
\fi
\hypersetup{
  pdfauthor   = {Constantin Loizides},
  pdftitle    = {},
  pdfcreator  = {},
  pdfproducer = {},
  pdfsubject  = {},
  pdfkeywords = {}
 }
\fi

\graphicspath{{./pics/}}

\usepackage{fancyhdr}
\usepackage{pslatex}
\def\PHOBOS{P\kern-.34em \lower.4ex\hbox{H}\kern-.12em \lower.6ex\hbox{O}\kern-.08em \lower-.18ex\hbox{B}\kern-.12em \lower-.45ex\hbox{O}\kern-.12em \lower.12ex\hbox{S}\ }
\newcommand {\snn}       {\ensuremath{\sqrt{s_{_{\rm NN}}}}}

\newcommand {\av}[1]     {\ensuremath{\left< #1 \right>}}
\newcommand {\abs}[1]    {\ensuremath{\left| #1 \right|}}

\newcommand {\hrefurl}[1]{\href{#1}{\url{#1}}}
\newcommand {\arxiv}[1]  {\href{http://www.arxiv.org/#1}{\mbox{arXiv:#1}}}
\newcommand {\Ref}[1]    {Ref.~\cite{#1}}
\newcommand {\Refs}[1]   {Refs.~\cite{#1}}
\newcommand {\eq}[1]     {eq.~(\ref{#1})}

\newcommand {\fig}[1]    {fig.~\ref{#1}}
\newcommand {\Fig}[1]    {Fig.~\ref{#1}}

\begin{document}

\title{Elliptic flow, eccentricity and eccentricity fluctuations}

\author{Constantin LOIZIDES$^{4}$ for the PHOBOS collaboration\\
\vspace{2mm}
%
%
B.Alver$^4$,
B.B.Back$^1$,
M.D.Baker$^2$,
M.Ballintijn$^4$,
D.S.Barton$^2$,
R.R.Betts$^6$,
R.Bindel$^7$,
W.Busza$^4$,
Z.Chai$^2$,
V.Chetluru$^6$,
E.Garc\'{\i}a$^6$,
T.Gburek$^3$,
K.Gulbrandsen$^4$,
J.Hamblen$^8$,
I.Harnarine$^6$,
C.Henderson$^4$,
D.J.Hofman$^6$,
R.S.Hollis$^6$,
R.Ho\l y\'{n}ski$^3$,
B.Holzman$^2$,
A.Iordanova$^6$,
J.L.Kane$^4$,
P.Kulinich$^4$,
C.M.Kuo$^5$,
W.Li$^4$,
W.T.Lin$^5$,
S.Manly$^8$,
A.C.Mignerey$^7$,
R.Nouicer$^2$,
A.Olszewski$^3$,
R.Pak$^2$,
C.Reed$^4$,
E.Richardson$^7$,
C.Roland$^4$,
G.Roland$^4$,
J.Sagerer$^6$,
I.Sedykh$^2$,
C.E.Smith$^6$,
M.A.Stankiewicz$^2$,
P.Steinberg$^2$,
G.S.F.Stephans$^4$,
A.Sukhanov$^2$,
A.Szostak$^2$,
M.B.Tonjes$^7$,
A.Trzupek$^3$,
G.J.van~Nieuwenhuizen$^4$,
S.S.Vaurynovich$^4$,
R.Verdier$^4$,
G.I.Veres$^4$,
P.Walters$^8$,
E.Wenger$^4$,
D.Willhelm$^7$,
F.L.H.Wolfs$^8$,
B.Wosiek$^3$,
K.Wo\'{z}niak$^3$,
S.Wyngaardt$^2$,
B.Wys\l ouch$^4$\\
\vspace{2mm}
\small
%
%
%
%
$^1$~Argonne National Laboratory, Argonne, IL 60439-4843, USA\\
$^2$~Brookhaven National Laboratory, Upton, NY 11973-5000, USA\\
$^3$~Institute of Nuclear Physics PAN, Krak\'{o}w, Poland\\
$^4$~Massachusetts Institute of Technology, Cambridge, MA 02139-4307, USA\\
$^5$~National Central University, Chung-Li, Taiwan\\
$^6$~University of Illinois at Chicago, Chicago, IL 60607-7059, USA\\
$^7$~University of Maryland, College Park, MD 20742, USA\\
$^8$~University of Rochester, Rochester, NY 14627, USA
\vspace{2mm}
}

\received{December 13,2006}

\begin{abstract}
Differential studies of elliptic flow are one of the most powerful tools
in studying the initial conditions and dynamical evolution of heavy ion
collisions. The comparison of data from Cu+Cu and Au+Au collisions taken
with the PHOBOS experiment at RHIC provides new information on the
interplay between initial geometry and initial particle density in
determining the observed final state flow pattern. 
Studies from PHOBOS point to the importance of fluctuations in the initial 
state geometry for understanding the Cu+Cu data. We relate the elliptic flow
data to the results of our model studies on initial state geometry fluctuations 
and discuss how we will perform measurements of event-by-event fluctuations in 
elliptic flow in Au+Au collisions. 
\keywords{Participant eccentricity, event-by-event}
\end{abstract}

\maketitle

\thispagestyle{fancy}

\setcounter{page}{1}

\section{Elliptic Flow}
Studies of the elliptic flow of produced particles via measurements of 
their azimuthal distribution are important probes of the dynamics 
of heavy ion collisions.
In the collision of two nuclei with finite impact parameter,
the almond-shaped overlap region has an azimuthal spatial
asymmetry. The asymmetrical shape of the source region can only 
be reflected in the azimuthal distribution of detected particles
if the particles significantly interact after their initial production.
Thus, observation of azimuthal anisotropy in the outgoing particles
is direct evidence of interactions between the produced particles. 
In addition, these interactions must occur at relatively early 
times, since expansion of the source ---even if uniform--- would gradually 
erase the magnitude of the spatial asymmetry. Typically, hydrodynamical 
models are used to quantitatively relate a specific initial source shape 
and the distribution of emitted particles~\cite{kolb2000}. 
At top RHIC energy, $\snn=200$~GeV, the elliptic flow signal at midrapidity 
in Au+Au collisions is described under the assumption of a boost-invariant 
relativistic hydrodynamic fluid, indicating that there is early equilibration 
in the colliding system~\cite{WhitePaper}.

The azimuthal anisotropy of the initial collision region can be characterized
by the eccentricity ($\varepsilon$) of the overlap region of the colliding nuclei 
in the transverse plane.
The strength of the elliptic flow, $v_2$, is commonly defined by the coefficient
of the second harmonic in the Fourier expansion of the azimuthal particle
distribution relative to the reaction plane, $\Psi_{\rm R}$, such that
$v_2 = \av{\cos(2\phi-2\Psi_{\rm R})}$~\cite{poskanzer98}.

Recent hydrodynamical calculations~\cite{bhalerao2005} conclude that $v_2$
scales approximately with $\varepsilon$ for small $\varepsilon$.
Such calculations typically use smooth, event-averaged initial conditions, for
which the initial azimuthal asymmetry is well described by the ``standard'' 
eccentricity,
$\varepsilon_{\rm std}
 =\frac{\sigma_{y}^2-\sigma_{x}^2}{\sigma_{x}^2+\sigma_{y}^2}$,
where $\sigma^2_{x}$~($\sigma^2_{y}$) is the variance of
the participant nucleon distribution projected on the $x$ ($y$) axis, taken
to be along (perpendicular to) the impact parameter direction.

\begin{figure}[t!]
 \centerline{
  \includegraphics[width=0.45\textwidth]{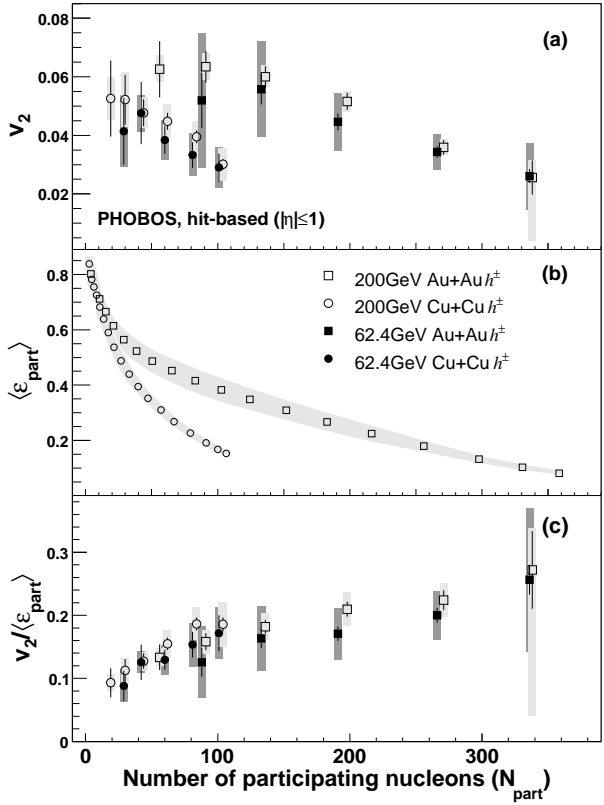}
 }
 \vspace{-0.2cm}
 \caption{
  (a) $v_2$ (unscaled), 
  (b) $\langle \varepsilon_{\rm part} \rangle$ and
  (c) $v_2$/$\langle \varepsilon_{\rm part} \rangle$ vs.\ N$_{\rm part}$ 
  for Cu+Cu and Au+Au collisions at $\snn=62.4$ and $200$~GeV.
  1-$\sigma$ statistical errors (bars) and 90\% C.L.\ systematic errors 
  (bands) are shown. Data points and eccentricity calculation are from
  \Refs{PhobosFlowPRL2,PhobosFlowPRL3}.\label{fig1}}
\end{figure}

PHOBOS has measured elliptic flow as a function of pseudorapidity, centrality, 
transverse momentum, center-of-mass 
energy~\cite{PhobosFlowPRL1,PhobosFlowPRL2,PhobosFlowPRC}
and, recently, nuclear species~\cite{PhobosFlowPRL3}.
In particular, the measurements of elliptic flow as a function of centrality
provide information on how the azimuthal anisotropy of the initial
collision region drives the azimuthal anisotropy in particle production. 
In \fig{fig1}a, we show the centrality dependence of $v_2$ at 
mid-rapidity~($\abs{\eta}<1$) for Cu+Cu and Au+Au at $\snn=$62.4 and 200~GeV collision 
energies, as obtained from our hit-based analysis 
method~\cite{PhobosFlowPRL2,PhobosFlowPRL3}.
A substantial flow signal is measured in Cu+Cu at both energies, even 
for the most central events. This is quite surprising, as according to 
the initial anisotropy given by $\varepsilon_{\rm std}$ one expects 
that $v_2$ should approach zero as the collisions become more central, 
similar to what has been found for the Au+Au case.

\section{Eccentricity fluctuations}
\enlargethispage{0.5cm}
As a possible explanation for the large $v_2$ signal in the small Cu+Cu system,
we have argued that event-by-event fluctuations in the shape of the
initial collision region may drive the elliptic flow~\cite{PhobosFlowPRL3,ManlyQM05}. 
For small systems or small transverse overlap regions, fluctuations 
in the nucleon positions frequently create a situation where the minor 
axis of the overlap ellipse 
is not aligned with the impact parameter vector. These fluctuations 
are neglected in the definition of $\varepsilon_{\rm std}$. Instead,
we have introduced the ``participant eccentricity'', 
\begin{equation}
 \varepsilon_{\rm part} = 
 \frac{\sqrt{(\sigma_{y}^2-\sigma_{x}^2)^2+4\sigma_{xy}^2}}{\sigma_{x}^2+\sigma_{y}^2}\,.
 \label{eqeccpart}
\end{equation} 
This definition accounts for the nucleon position fluctuations by quantifying the 
eccentricity event-by-event with respect to the minor axis of the overlap ellipse,
in the frame, defined by $\Psi_{\rm part}$, that diagonalizes the ellipse. Note,
$\sigma_{xy}=\langle xy\rangle - \langle x\rangle\langle y\rangle$, 
$\sigma^2_{x}$ and $\sigma^2_{y}$ are the (co-)variances of the $x$ and $y$ participant 
nucleon position distributions expressed in the original frame, given by $\Psi_{\rm R}$.
For a system with a large number of nucleons, the covariance term is comparatively small. 
Therefore, the average values of $\varepsilon_{\rm std}$ and $\varepsilon_{\rm part}$ over
many events are similar for all, but the most peripheral interactions for the Au+Au 
system. For the smaller Cu+Cu system, however, fluctuations in the nucleon positions 
become more important for all centralities~\cite{PhobosFlowPRL3}.

In \fig{fig1}b, we show a Glauber model calculation of $\varepsilon_{\rm part}$ as a 
function of $N_{\rm part}$ for Cu+Cu and Au+Au. The colliding nuclei are built by randomly 
placing nucleons according to a Woods-Saxon distribution. Excluded volume effects are
addressed by requiring a minimum inter-nucleon separation distance of $0.4$~fm. 
As opposed to $\varepsilon_{\rm std}$, where averages are implicitly over participants and 
events, the variance expressions in $\varepsilon_{\rm part}$ are averaged event-by-event 
over participants, individually. To check how the event-by-event interpretation of the 
Glauber calculation depends on the external parameter settings, we varied a number of 
sources of systematic error, like the nuclear radius, nuclear skin depth, nucleon-nucleon 
inelastic cross-section $\sigma_{\rm NN}$ and minimum nucleon separation. Varying each 
specific parameter within reasonable limits, the individual contributions were added in 
quadrature to determine the 90\% confidence level errors shown in \fig{fig1}b.

In order to compare the elliptic flow signals across nuclear species and with hydrodynamical 
predictions, it is important to scale out the difference in the initial asymmetry of the 
collision geometry. In \fig{fig1}c, we show the eccentricity-scaled flow for Cu+Cu and 
Au+Au, $v_2/\langle \varepsilon_{part} \rangle$, as a function of centrality. The scaled 
data are very similar for both the Cu+Cu and Au+Au collision systems at the 
same number of participants.
It is important to note that the apparent scaling does 
not rely on the fine-tuning of the Glauber parameter settings, as is evident from the systematic 
errors, which are rather small (see \fig{fig1}b). 

\section{Elliptic flow fluctuations}
If the proposed participant fluctuations are driving the value of $v_2$ event-by-event,
as suggested by our $\langle v_2 \rangle$ analysis, then we expect them to contribute 
to event-by-event elliptic flow fluctuations.
As mentioned above, ideal hydrodynamics leads to $v_2\propto\varepsilon$~\cite{bhalerao2005}. 
Assuming the same relation holds event-by-event, this would imply that
${\sigma_{v_2}}/{\langle v_2 \rangle} = {\sigma_{\varepsilon}}/{\langle \varepsilon \rangle}$,
where $\sigma_{v_2}$ ($\sigma_{\varepsilon}$) is the standard deviation of the event-by-event distribution 
of $v_2$ ($\varepsilon$). Using our $\langle v_2\rangle$ data (\fig{fig1}a) and the Monte Carlo Glauber
simulation to obtain $\sigma_{\varepsilon}$ (and $\langle \varepsilon \rangle$) for the participant
eccentricity, we estimate $\sigma_{v_2}$ to be about $2$\% for all centralities, except the most central
Au+Au collisions at $\snn=200$~GeV. This estimate leads to rather large relative 
fluctuations~($\sigma_{v_2}/\langle v_2 \rangle$) between $35$ and $50$\%. It is important 
to note that these estimates neglect other sources of elliptic flow fluctuations.

\begin{figure}[t!]
 \centerline{
  \includegraphics[width=0.5\textwidth]{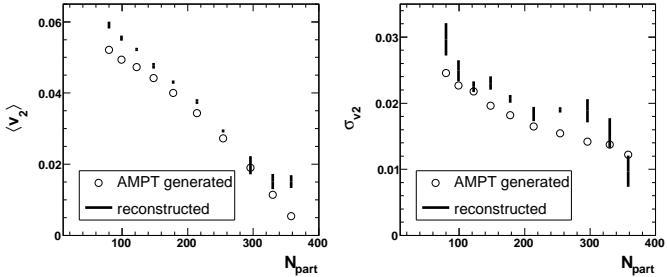}
 }
 \vspace{-0.2cm}
 \caption{
  Generated and reconstructed $\langle v_2 \rangle$~(left) and $\sigma_{v_{2}}$ for AMPT
  vs.\ $N_{\rm part}$. The generated signal is extracted from a mixed-event analysis based
  on the true Monte Carlo particle information. The reconstructed signal is averaged over
  results of running the complete flow fluctuation analysis in 10 vertex bins using the 
  kernel created from the modified HIJING samples. The black lines represent the $\pm1\sigma$ 
  error on the mean, with the mean centered (not shown).
\label{fig2}}
\end{figure}

In the following, we will give a brief overview of the analysis method
that we have developed to measure flow fluctuations with PHOBOS in peripheral and
semi-central Au+Au collisions. Details can be found in \Ref{Alver06}.
We seek to discriminate known (statistical)
from unknown (presumably dynamical) contributions to observed flow fluctuations.
Ideally, they would add according to
\mbox{$\sigma_{v_{2},\text{obs}}^2 = \sigma_{v_{2},\text{dyn}}^2 + \sigma_{v_{2},\text{stat}}^2$}.
This relation holds if the average of the measurement, $\langle
v_{2}^{\text{obs}}\rangle$, gives the true average in the data, $\langle
v_{2}\rangle$, and if the resolution of our method is independent of
the true value. Neither of these conditions are fully satisfied in the
event-by-event measurement of $v_2$. Therefore, a more detailed knowledge
of the response function is required. 
We define $K(v_{2}^{\text{obs}},v_{2}, n)$ as the distribution of the
event-by-event observed elliptic flow, $v_{2}^{\text{obs}}$, for events with
constant input flow value, $v_2$, and multiplicity, $n$. Assuming a set of events 
has an input $v_2$ distribution given by $f(v_2)$, then $g(v_{2}^{\text{obs}})$, 
the distribution of $v_{2}^{\text{obs}}$, will be given according to
\begin{equation}
 g(v_{2}^{obs}) = \int^{\infty}_{0} K(v_{2}^{\text{obs}},v_2,n) \, f(v_2) 
 \, N(n) \, \text{d}v_2  \, \text{d}n\,,
 \label{eqkernel}
\end{equation}
where $N(n)$ is the multiplicity distribution of the events in the given set of 
events~(centrality bin).
Thus, our event-by-event elliptic flow fluctuation analysis consists of 
three steps:
\begin{itemize}
\item Finding the observed $v_{2}$ distribution, $g(v_{2}^{\text{obs}})$, for a set 
of events by an event-by-event measurement of $v_{2}^{\text{obs}}$.
\item Construction of the kernel, $K(v_{2}^{\text{obs}},v_2,n)$, by studying the
detector response for sets of constant (known) input value of $v_2$ and 
multiplicity $n$. 
\item Calculating the true $v_2$ distribution, $f(v_2)$, by finding a
solution to \eq{eqkernel}.
\end{itemize}

For the event-by-event measurement we use a maximum likelihood method, making use of 
all hit information from the multiplicity array to measure a single value, $v_{2}^{obs}$, 
while allowing an efficient correction for the non-uniformities in the acceptance.
We model the measured pseudorapidity dependence of $v_2$ according to 
$v_2(\eta)=v_2\cdot (1-\abs{\eta}/6)$, with $v_2\equiv v_2(0)$, which is known
to describe our mean $v_2$ data reasonable well.
We define the probability distribution function (PDF) of a
particle to be emitted in the direction $(\eta,\phi)$ for an event with $v_{2}$ 
and reaction plane angle $\phi_0$ as
$P(\eta,\phi|v_2,\phi_{0}) = p(\eta)[1+2v_{2}(\eta)\cos(2\phi-2\phi_{0})]$.
The normalization $p(\eta)$ is constructed such that the PDF, folded with the PHOBOS
acceptance, yields the same value for different sets of parameters~($v_{2}$,$\phi_0$). 
Maximizing $\prod_{i=1}^{n} P(\eta_i,\phi_i|v_2,\phi_{0})$ as a function of 
$v_{2}$ and $\phi_0$ allows us to measure $v_{2}^{\text{obs}}$ event-by-event.
We determine the response function $K(v_2^{\text{obs}},v_2,n)$ in bins
of $v_2$ and $n$ using modified HIJING \ifcomment~\cite{HIJING}\fi events. 
Flow of constant magnitude ($v_{2}$) with a flat reaction plane distribution ($\phi_0$)
is introduced into generated HIJING Au+Au events. This is achieved by redistributing 
the resulting particles in each event in $\phi$ randomly according to 
$1+2v_{2}(\eta)\cos(2\phi-2\phi_{0})$, using their generated $\eta$ positions.
The modified HIJING events are run through GEANT \ifcomment~\cite{GEANT}\fi to obtain 
the PHOBOS detector response.
To finally extract $f(v_2)$, we for now simply assume a Gaussian distribution with
$\langle v_{2}\rangle$ and $\sigma_{v_2}$. For given values of $\langle v_{2}\rangle$ 
and $\sigma_{v_2}$, it is possible to take the integral in \eq{eqkernel} to obtain the 
expected distribution, \mbox{$g_{\text{exp}}(v_2^{\text{obs}}|\langle
v_{2}\rangle,\sigma_{v_2})$}. By comparing the expected and observed distributions, 
values for $\langle v_2 \rangle$ and $\sigma_{v_2}$ are obtained from a minimized $\chi^2$ 
of the data to the expectation.
As outlined in \Ref{Alver06}, the whole analysis procedure was verified
on similar HIJING events as used to construct the kernel, and 
found to successfully reconstruct
the input fluctuations, provided $\langle v_2\rangle \ge 0.03$. 

Here, we report on a study with fully simulated AMPT \ifcomment~\cite{AMPT}\fi events 
in order to verify the analysis with a different type of ``data'' events than 
used to create the kernel, since in AMPT flow builds up dynamically. 
To minimize acceptance effects, the analysis is done in bins of collision 
vertex. \Fig{fig2} shows the averaged results obtained from 
the different vertex bins compared to the generated signal. 
Since the information about the generated 
$v_2$ in AMPT has not been readily available on an event-by-event basis, we 
extracted the generated $\av{v_2}$ and $\sigma_{v_2}$ from a mixed-event 
analysis based on the true Monte Carlo particle information. 
We conclude
that the developed analysis chain is able to reconstruct the fluctuations
to a satisfactory degree over a large range in centrality, in samples
of ``data'' that are different from the ones used to construct the kernel.


%
%
%
%
This work was partially supported by U.S. DOE grants
DE-AC02-98CH10886,
DE-FG02-93ER40802,
DE-FC02-94ER40818,  
DE-FG02-94ER40865,
DE-FG02-99ER41099, and
W-31-109-ENG-38, by U.S.
NSF grants 9603486, 
0072204,            
and 0245011,        
by Polish KBN grant 1-P03B-062-27(2004-2007),
by NSC of Taiwan Contract NSC 89-2112-M-008-024, and
by Hungarian OTKA grant (F 049823).

\end{document}